\documentclass{ws-procs10x7}
\usepackage{balance}

\newcolumntype{d}[1]{D{.}{.}{#1}}

\newcommand{\notD}{\ \not\!\!{D}}
\newcommand{\notDbar}{\ \not\!\!{\overline{D}}}
\newcommand{\etal}{{\em et al.}}
\newcommand\babar{\mbox{\textsc{BaBar}}}

\makeindex
\begin{document}

\title{Status of $|V_{ub}|$, $|V_{cb}|$ and their relative phase}

\author{R.~V.~Kowalewski$^*$}

\address{Department of Physics and Astronomy, University of Victoria, 
Victoria, BC V8N 2X3, Canada\\$^*$E-mail: kowalews@uvic.ca}

\twocolumn[\maketitle\abstract{The current status
of the determinations of the CKM elements $V_{ub}$
and $V_{cb}$ is reviewed and future prospects are discussed.  }
\keywords{CKM matrix; CP violation; B physics.}
]

\section{Motivation}

The imaginary phase in the Cabibbo-Kobayashi-Maskawa (CKM) quark
mixing matrix has, in recent years, been firmly established as the
dominant source of CP violation in the decays of $B$ mesons.  In the
process, constraints on the lesser-known parameters of the
CKM matrix, namely $\bar{\rho}$ and $\bar{\eta}$,\footnote{The
parameterization used here has $\bar{\rho}=(1-\lambda^2/2)\rho$,
$\bar{\eta}=(1-\lambda^2/2)\eta$, and defines $V_{us}=\lambda$,
$V_{cb}=A\lambda^2$ and $V_{ub}=A\lambda^3(\rho-i\eta)$.}  have become
increasingly precise.  Measurements of CP asymmetries determine the
angles of the unitarity triangle in the $\bar{\rho}$-$\bar{\eta}$
plane.  These all involve processes with internal loops, either
through Penguin amplitudes or through $B^0\overline{B}{}^0$ mixing.
These same CKM parameters can be determined in tree-level processes,
which are essentially immune to contributions from new physics, by
measuring the magnitudes $|V_{ub}|$ and $|V_{cb}|$ and the relative
phase $\gamma=\phi_3\equiv arg[-(V_{ud}V^*_{ub})/(V_{cd}V^*_{cb})]$.
The independent determination of $\overline{\rho}$ and
$\overline{\eta}$ in tree and loop-dominated processes
thus provides a promising avenue in which to search for
deviations from the Standard Model.  This talk summarizes the
current status of determinations of $|V_{ub}|$, $|V_{cb}|$ and their
relative phase, and discusses prospects for the near-term improvement of
these measurements.

The long $B$ lifetime\cite{Blife} determined by MAC and Mark-II, along
with the prevelance of charm mesons in $B$ decays, indicated that
$|V_{cb}|$ was small and that $|V_{ub}|$ was smaller still.  The first
indications of a non-zero $|V_{ub}|$ came from CLEO and ARGUS in
1990.\cite{firstVub} Increasingly precise determinations of these
magnitudes have been made ever since, and the uncertainty on
$|V_{cb}|$ is now only three times that of $\lambda$, the best-known
CKM parameter.

Due to space limitations, only recent developments will be discussed
and cited in this review.  More comprehensive reviews of $|V_{ub}|$
and $|V_{cb}|$\cite{PDGreview} and of deteminations of $\gamma\
(\phi_3)$\cite{gammaReview} are available.

\section{Semileptonic $B$ decays}

The presence of a single hadronic current renders the semileptonic
decay width calculable with modest theoretical uncertainties, and
makes these decays the favored system for determinations of the
magnitudes $|V_{ub}|$ and $|V_{cb}|$.  Nevertheless, uncertainties
related to non-perturbative QCD comprise a significant part of the
total uncertainty in these determinations.  The theoretical methods
used to calculate the decay rates as a function of $|V_{qb}|$
are very different for inclusive and exclusive decays, as are the
experimental measurements, allowing the comparison of these complementary
determinations to provide an important check on the results.

Measurements of $|V_{ub}|$ and $|V_{cb}|$ are dominated by the
experiments at the $\Upsilon{(4S)}$ resonance,
namely Belle, \babar\ and CLEO.  These experiments measure $B$ mesons
produced nearly at rest in the center-of-mass frame.  These
$B\overline{B}$ events result in nearly isotropic distributions of
final state particles, and are separated from the more collimated
$e^+e^-\to q\overline{q}$ ($q=d,u,s,c$) interactions using event shape
cuts.  The residual $q\overline{q}$ background is determined using
data collected 
just below the $B\overline{B}$ production threshold.

\subsection{Inclusive semileptonic decays}

In inclusive semileptonic decays, a subset of the final state
particles (e.g. just the lepton) are identified and measured,
integrating over all decay channels and the kinematics of unmeasured
particles, resulting in singly- or doubly-differential partial widths.
Low-order moments of these distributions are measured and compared
with theoretical calculations to determine $|V_{qb}|$, the $b$ quark
mass, and related non-perturbative parameters.

\subsubsection{Theoretical framework}
The fact that the $b$ quark mass $m_b$ is large compared to
$\Lambda_{QCD}$ allows for a separation of scales as the basis of an
effective field theory, the Heavy Quark Expansion (HQE).  In the HQE
the short-distance degrees of freedom are integrated out, resulting in
a double expansion in powers of $\Lambda_{QCD}/m_b$ and of
$\alpha_S(\mu)$, with $\mu\gg \Lambda_{QCD}$.  The expression for the
total semileptonic $b\to c\ell\overline{\nu}$ decay width is given in
Eq.~\ref{eq-GammaVcb},
\begin{table*}[bt]
 \begin{eqnarray}
\label{eq-GammaVcb}
  &&  \Gamma = |V_{cb}|^2 \frac{G_F^2 m_b^5(\mu)}{192 \pi^3}
               (1+A_{\mathrm{ew}}) A^{\rm pert} (r,\mu) \times
    \\  \nonumber&&  \left[z_0 (r) + z_1(r)\times 0 +
     z_2 \left(r, \frac{\mu_\pi^2}{m_b^2}, \frac{\mu_G^2}{m_b^2}\right)
   + z_3 \left(r, \frac{\rho_{\rm D}^3}{m_b^3} , \frac{\rho_{\rm LS}^3}{m_b^3}\right)
    + ... \right]    \,
\end{eqnarray}   
\end{table*}
where $A_{\mathrm{ew}}$ and $A^{\mathrm{pert}}(r,\mu)$ denote the electroweak 
and QCD perturbative corrections and $r=m_c/m_b$.  The $z_i$ are known
functions and depend on non-perturbative
parameters ($\mu_\pi^2$, etc.) that correspond to matrix
elements of local operators divided by the appropriate power of $m_b$.  
The coefficient of the $1^{\mathrm{st}}$-order term vanishes, so the 
leading corrections are at the percent level.  Similar expressions,
involving the same non-perturbative parameters, have been calculated
for low-order moments of the lepton momentum and squared hadron mass 
spectra  in $b\to c\ell\overline{\nu}$ decays, as well as for 
$b\to u\ell\overline{\nu}$ decays and for the inclusive radiative 
decay $b\to s\gamma$.

In all cases, comparison of these calculated inclusive decay rates
with measured rates depend on the assumption of quark-hadron duality.
It's clear that this assumption breaks down in restricted regions of
phase space (e.g. at low squared hadronic mass, where only discrete
values are physically realized).  This is the motivation for
concentrating on low-order moments of decay spectra, 
integrated over broad regions of phase space.  While the
uncertainty due to this assumption remains hard to quantify, the
global fit to a large number of spectral moments using a small
number of parameters, discussed in the following section, suggests
that any violations are small compared to the current level of
sensitivity.

\subsubsection{Determination of $\boldmath{|V_{cb}|}$}
The theoretical calculations of low-order moments of the $b\to
c\ell\overline{\nu}$ and $b\to s\gamma$ decay spectra have been
performed, for a variety of requirements on the minimum lepton
momentum or photon energy, in two separate mass renormalization
schemes, referred to here as the ``kinetic''\cite{ref-kinetic} and
``1S''\cite{ref-1S} schemes.  Each calculated moment depends on the
quark masses and on a common set of non-perturbative parameters.  The
total $b\to c\ell\overline{\nu}$ rate also depends on $|V_{cb}|^2$.

A variety of experiments have measured moments of these decay
processes.  Measurements of the lepton momentum
moments\cite{lepton-moments} are based on a technique introduced by
\mbox{ARGUS,\cite{argus-dilepton}} in which charge and angular correlations
in events with two identified leptons are used to extract the direct
$B\to X\ell\overline{\nu}$ spectrum down to $\sim 0.5\,$GeV.  These
measurements are limited by systematic uncertainties, although some
further improvement may be possible.  Measurements of the moments of
the squared hadronic mass spectrum,\cite{hadron-mass-moments} for
various cuts on the minimum lepton momentum, have been made.  The most
precise of these rely on fully reconstructing one $B$ meson from
$\Upsilon{(4S)}$ decay to allow the remaining particles in the event
to be associated with the semileptonic decay of the second $B$ meson.
Measurements of the photon energy spectrum from $b\to s\gamma$
decays\cite{photon-moments} are experimentally challenging, in
particular as the minimum accepted photon energy is reduced below
about $2.2\,$GeV, due to the large background from both
$q\overline{q}$ and $B\overline{B}$ events.  

The measured moments, including all known correlations, have been
fitted\cite{Buchmuller} in the kinetic scheme, resulting in
precise values for $m_b$ and $|V_{cb}|$, as shown in
Table~\ref{tab-inclusiveVcb}.  Separate fits to the $b\to
c\ell\overline{\nu}$ moments alone and to the $b\to s\gamma$ moments
(with only $m_b$ and $\mu_\pi^2$ floating) give consistent results.
The fit includes both experimental and theoretical uncertainties, and
results in a $\chi^2$ of 19.3 for 44 degrees of freedom.  The first
and second uncertainties come from experimental errors and
uncertainties in the HQE calculations, respectively.  The last
uncertainty listed for $|V_{cb}|$ is a normalization uncertainty due
to uncalculated terms in the total rate.

The global fit described above does not yet include the latest moment
measurements from Belle, which were first presented at this
conference.\cite{BelleMomentsFit} Belle performs a fit to their
measured $b\to c\ell\overline{\nu}$ and $b\to s\gamma$ moments in both
the kintetic and 1S schemes, resulting in $\chi^2/$d.o.f.  of 17.8/24
and 5.7/17, respectively, and in the values given in
Table~\ref{tab-inclusiveVcb}.  For the 1S fit the first error listed
includes both experimental and theoretical uncertainties; the second
error on $|V_{cb}|$ comes from the $B$ meson lifetime.  The values for
$|V_{cb}|$ agree well in all three fits.  The $m_b$ values in the
kinetic and 1S schemes cannot be compared directly; when both are
translated to a common scheme the agreement is excellent.

\begin{table}[htb]
\tbl{Fitted values for $|V_{cb}|$ and $m_b$.  The fits are described
in the text. \label{tab-inclusiveVcb}}
{\begin{tabular}{@{}lcc@{}}
\toprule
Fit            & $|V_{cb}|\ (10^{-5})$ & $m_b$ (MeV)\\ \colrule
Global kin. & $4196\pm 23\pm 35\pm 59$ & $4590\pm 25\pm 30$ \\
Belle kin.  & $4206\pm 67\pm 48\pm 63$ & $4564\pm 76$ \\
Belle 1S       & $4149\pm 52\pm 20$       & $4729\pm 48$ \\
\botrule
\end{tabular}}
\end{table}

These precise determinations of $|V_{cb}|$ and $m_b$ in a
consistent global fit represent an enormous achievement.

\subsubsection{Determination of $\boldmath{|V_{ub}|}$}

The selection of events of the type $b\to u\ell\overline{\nu}$
requires suppression of the dominant ($\times 50$) background from the
process $b\to c\ell\overline{\nu}$.  Kinematic criteria on the lepton
momentum, on the squared momentum transfer ($q^2$) in the $b$ decay,
or on the invariant mass of the final state hadrons can be used to
reduce the background.  The dependence of the partial rate on $m_b$ in
the restricted phase space region is steeper than for the total rate;
typical values are $m_b^7$-$m_b^{12}$, depending on the experimental
cuts.  Restrictive kinematic cuts can compromise the
convergence of the HQE, in which case the calculated rate becomes
sensitive to the non-perturbative light-cone momentum distribution
(shape function), which, at leading order, must be measured (in the
radiative decays $b\to s\gamma$) or modelled.  Additional shape
functions, which differ in semileptonic and radiative decays, arise at
higher orders and must be modelled.

Significant improvements in the calculational methods have recently
become available.\cite{BLNP,DGE}  Calculations that relate directly
integrals over the measured lepton momentum or hadron invariant mass
spectra in $b\to u\ell\overline{\nu}$ decays with integrals over the
measured $E_\gamma$ spectrum in $b\to s\gamma$ decays are
available\cite{LLR,bulnu-bsg} and obviate the need to model the
leading shape function.

A recent \babar\ measurement\cite{BaBarSFfree} 
using the invariant mass $m_X$ of the
recoiling hadrons to select $b\to u\ell\overline{\nu}$ decays used a
sample of $88\,$million $B\overline{B}$ events to determine $|V_{ub}|$
in two separate ways.  The first method compared the integrated $m_X$
and $E_\gamma$ spectra using the calculations of Ref.~\refcite{LLR} to
determine $|V_{ub}|=(4.43\pm 0.38\pm 0.25\pm 0.29)\cdot 10^{-3}$,
where the errors are statistical, systematic and theoretical,
respectively.  The second method determined the inclusive $b\to
u\ell\overline{\nu}$ rate for $m_X<2.5\,$GeV, which includes 96\%\ of
the total rate, and results in $|V_{ub}|=(3.84\pm 0.70\pm 0.30\pm
0.10)\cdot 10^{-3}$.  While the impact of these measurements on the
world average $|V_{ub}|$ is small, they are noteworthy for the small
theoretical uncertainties, and will improve markedly as more data are
analyzed.

CLEO has produced the first direct limits\cite{CLEO-WA} on the uncertainty in
$|V_{ub}|$ determinations arising from weak annihilation diagrams,
which affect charged $B$ decays to an isoscalar hadron, lepton and neutrino.
These limits are used to evaluate the uncertainty in the world average
$|V_{ub}|$ from weak annihilation.

The Heavy Flavor Averaging Group (HFAG)\cite{HFAG-semi} provides
determinations of $|V_{ub}|$ based on a variety of
measurements.\cite{inclusiveVub}  The resulting average is
\begin{equation}
|V_{ub}|=(4.49\pm 0.19\pm 0.27)\cdot 10^{-3}
\end{equation}
based on the calculations of Ref.~\refcite{BLNP}.  The average has a
$\chi^2$ probability of 41\%.  The error budget consists of $2.2\%$
from statistics, $2.8\%$ from experimental systematics, $1.9\%$ from
weak annihilation, $1.9\%$ from the modeling of $b\to
c\ell\overline{\nu}$ decays, $1.6\%$ from the modeling of $b\to
u\ell\overline{\nu}$ decays, $3.8\%$ from the modeling of sub-leading
shape functions and perturbative matching scales, and $4.2\%$ from HQE
parameter uncertainties, principally from $m_b$.  The result using the
calculations of Ref.~\refcite{DGE} is $|V_{ub}|=(4.46\pm 0.20\pm
0.20)\times 10^{-3}$, with a $\chi^2$ probability of 12\%.  An
additional calculation\cite{BLL} is available for the subset of
measurements made with requirements on $m_X$ and $q^2$, and gives
$|V_{ub}|=(5.02\pm 0.26\pm 0.37)\times 10^{-3}$;
this value is compatible with those obtained using the other
calculations.  Many of the measurements in the average use relatively
small samples compared with the current and projected $B$-factory
datasets.  A 5\%\ uncertainty on $|V_{ub}|$ is an aggressive target
for the next ICHEP conference.

\subsection{Exclusive semileptonic decays}
Exclusive semileptonic decays provide a complementary avenue for
determinations of $|V_{qb}|$.  The challenge for theory is the
calculation of the decay form factor, in particular of its
normalization.

\subsubsection{$\boldmath{b\to c\ell\overline{\nu}}$ decays}
These decays involve a heavy-to-heavy transition, allowing
heavy quark symmetry to be applied.  In the heavy quark
limit this results in a unique form factor, the Isgur-Wise
function, that needs to be measured.  This form factor is
parameterized as a function of the four-velocity product $w$ of the
$B$ and charm mesons, which is related to the momentum
transfer $q^2$: $w=(m_B^2+m_C^2-q^2)/(2m_B m_C)$.

The decay $\overline{B}{}^0\to D^{*+}\ell^-\overline{\nu}$ is the easiest to
isolate experimentally, as it has the largest branching fraction of
any $B$ decay.  Many experiments have measured this decay mode,
extracting the decay rate versus $w$ to determine
$\mathcal{F}(1)|V_{cb}|$, where $\mathcal{F}(1)$ is the form factor
normalization at the zero recoil point, $w=1$.  Recently, \babar\
has improved measurements of the form factor slope
$\rho^2=-d\mathcal{F}/dw (w=1)$ and the form factor ratios
$R_1\sim V/A_1$ and $R_2\sim A_2/A_1$.  They find\cite{DstlnuFF}
\begin{eqnarray}
\rho^2&=&1.179\pm 0.048\pm 0.028\\
R_1&=&1.417\pm 0.061\pm 0.044\\
R_2&=&0.836\pm 0.037\pm 0.022
\end{eqnarray}
Since existing measurements of $\mathcal{F}(1)|V_{cb}|$ depend on
$R_1$ and $R_2$, these more precise values result in smaller
uncertainties on $|V_{cb}|$.  The HFAG average\cite{HFAG-semi} is
$\mathcal{F}(1)|V_{cb}|=(36.2\pm 0.8)\cdot 10^{-3}$.  Using
$\mathcal{F}(1)=0.919\,^{+0.030}_{-0.035}$ from Ref.~\refcite{F1-LQCD}
gives 
\begin{equation}
|V_{cb}|=(39.4\pm 0.9\,^{+1.6}_{-1.2})\cdot 10^{-3},
\end{equation}
which is consistent with the inclusive results quoted earlier.
Further progress is needed to reduce the uncertainty on the
calculation of the form factor, and the experimental situation needs
to be clarified, given the poor $\chi^2/$d.o.f., 38.7/14, of the
existing measurements.

It has been argued recently\cite{DlnuFF} that
the theoretical uncertainty on the form factor normalization for
$\overline{B}\to D\ell\overline{\nu}$ decays may be even smaller
than for $\overline{B}\to D^*\ell\overline{\nu}$ decays; this may
be true for lattice QCD determinations as well.  The precise
determination of the $\overline{B}\to D\ell\overline{\nu}$ form factor
at zero recoil remains an experimental challenge due to the large
background from $\overline{B}\to D^*\ell\overline{\nu}$ decays.

\subsubsection{$\boldmath{b\to u\ell\overline{\nu}}$ decays}

\begin{table}[htb]
\tbl{Recent measurements of exclusive
$\overline{B}\to \pi\ell\overline{\nu}$ decays.  The values for the
$B^+$ mode are multiplied by $2\tau(B^0)/\tau(B^+)$ to allow direct
comparison with the $B^0$ mode.
 \label{tab-pilnu}}
{\begin{tabular}{@{}lcc@{}}
\toprule
Expt/tag & $\mathcal{B}(B^0)\ (10^{-6})$ & $\mathcal{B}(B^+)\ (10^{-6})$\\ \colrule
\babar/none\cite{BaBar-pilnu-untagged} & $144\pm\ \,8\pm 10$ &\\
CLEO/none\cite{CLEO-pilnu} &  $137\pm 16\pm 13$ &\\
Belle/s.l.\cite{Belle-pilnu-tagged} & $138\pm 19\pm 14$ & $143\pm 26\pm 16$\\
Belle/had\cite{Belle-pilnu-tagged} & $149\pm 26\pm \ \,6$ & $160\pm 32\pm 11$\\
\babar/s.l.\cite{BaBar-pilnu-tagged} & $112\pm 25\pm 10$ & $135\pm 33\pm 19$\\
\babar/had\cite{BaBar-pilnu-tagged} & $107\pm 27\pm 19$ & $152\pm 41\pm 20$\\
\botrule
\end{tabular}}
\end{table}

\begin{figure}[b]
\centerline{\psfig{file=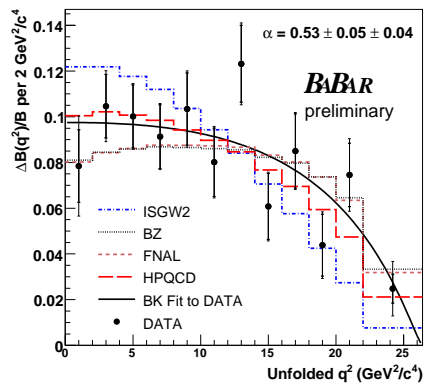,width=2.2in}}
\caption{Partial branching fraction versus $q^2$ for 
$\overline{B}{}^0\to\pi^+\ell^-\overline{\nu}$.}
\label{fig-pilnu}
\end{figure}

New measurements of the branching fraction versus $q^2$ for
$\overline{B}\to \pi\ell\overline{\nu}$ have significantly reduced
the experimental uncertainty in the determination of $|V_{ub}|$ from
these decays.  The measurements are done either with (tagged) or
without (untagged) the reconstruction of the other $B$ meson in the
event.  The tagged measurements provide superior signal to background
and resolution on $q^2$, but are less statistically precise than
untagged measurements.  Belle,\cite{Belle-pilnu-tagged} \babar\cite{BaBar-pilnu-untagged,BaBar-pilnu-tagged} and CLEO\cite{CLEO-pilnu} presented new
measurements for this conference.  Belle and \babar\ both provided
measurements of $\overline{B}\to\pi\ell\overline{\nu}$ with the other
$B$ reconstructed in the decay $B\to\overline{D}{}^{(*)}\ell^+\nu$ or
in an hadronic decay mode.  CLEO and \babar\ made untagged measurements
of $\pi\ell\overline{\nu}$ made based on neutrino reconstruction.  The
\babar\ measurement has very high statistics and provides a good
determination of the $q^2$ dependence (see Fig.~\ref{fig-pilnu}),
obtaining a shape parameter $\alpha=0.53\pm 0.05\pm 0.04$ for the
Becirevic-Kaidalov parameterization.\cite{Becirevic}  These recent
measurements are summarized in Table~\ref{tab-pilnu}.  
Measurements of
$B\to\rho\ell\nu$,\cite{Belle-pilnu-tagged,CLEO-pilnu} and of
$B\to\eta\ell\nu$ and
$B\to\eta^\prime\ell\nu$\cite{BaBar-etalnu-tagged} were also presented
at the conference.

HFAG has averaged all available $\overline{B}\to\pi\ell\nu$
measurements and finds, assuming isospin symmetry for the decay rates,
$\mathcal{B}(B^0\to \pi^-\ell^+\nu)=(1.37\pm 0.06\pm 0.06)\cdot
10^{-4}$.  The consistency of the measurements is good.  Comparing the
partial decay rates in the region $q^2>16\,$GeV$^2$ with calculations
of the form factor normalization from Lattice QCD, and those in the
region $q^2<16\,$GeV$^2$ with calculations from light-cone sum rules,
results in the $|V_{ub}|$ values in Table~\ref{tab-Vubpilnu}.

\begin{table}[htb]
\tbl{\label{tab-Vubpilnu}
$|V_{ub}|$ from $\overline{B}\to\pi\ell\overline{\nu}$.}
{\begin{tabular}{@{}lc@{}}
\toprule
FF calculation         & $|V_{ub}|$ $(10^{-3})$ \\ \hline
Ball/Zwicky\cite{LCSR} & $3.38\pm 0.12\,{}^{+0.56}_{-0.37}$ \\
HPQCD\cite{HPQCD}      & $3.93\pm 0.26\,{}^{+0.59}_{-0.41}$ \\
FNAL\cite{FNAL}        & $3.51\pm 0.23\,{}^{+0.61}_{-0.40}$ \\
APE\cite{APE}          & $3.54\pm 0.23\,{}^{+1.36}_{-0.63}$ \\
\botrule
\end{tabular}}
\end{table}

\noindent The experimental uncertainties are already at the $\sim 6\%$-level.
Progress is clearly needed in the form factor normalization
calculations to provide a competitive determination of $|V_{ub}|$.
These values are not independent, and are lower than the 
inclusive determination of $|V_{ub}|$ by $0.7$-$1.7\sigma$.

\section{The relative phase $\boldmath{\gamma\ (\phi_3)}$}

\begin{figure}[htb]
\centerline{\psfig{file=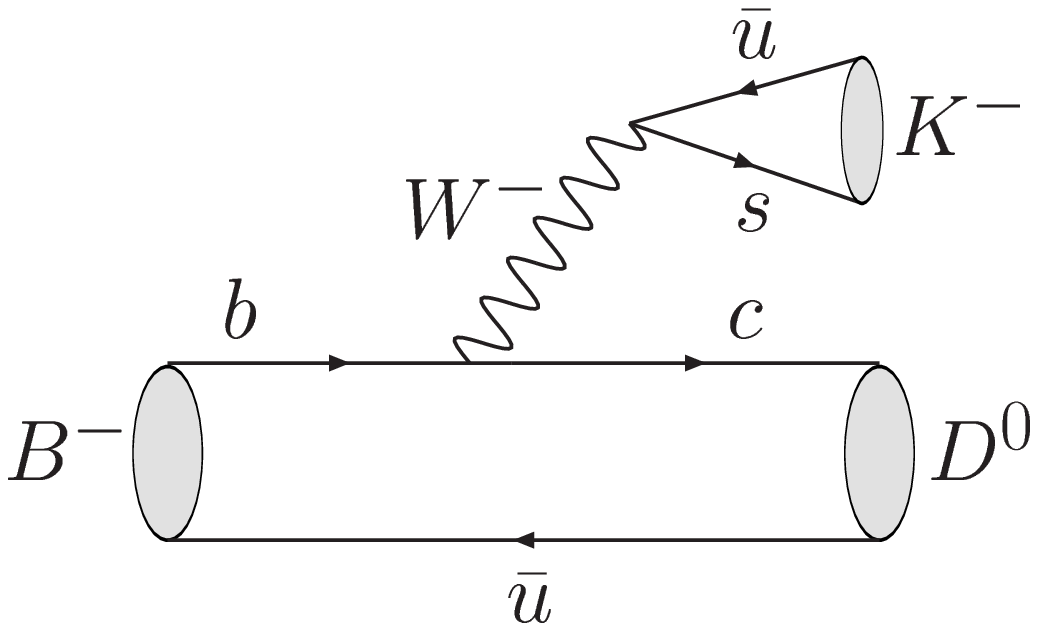,width=1.3in}
\psfig{file=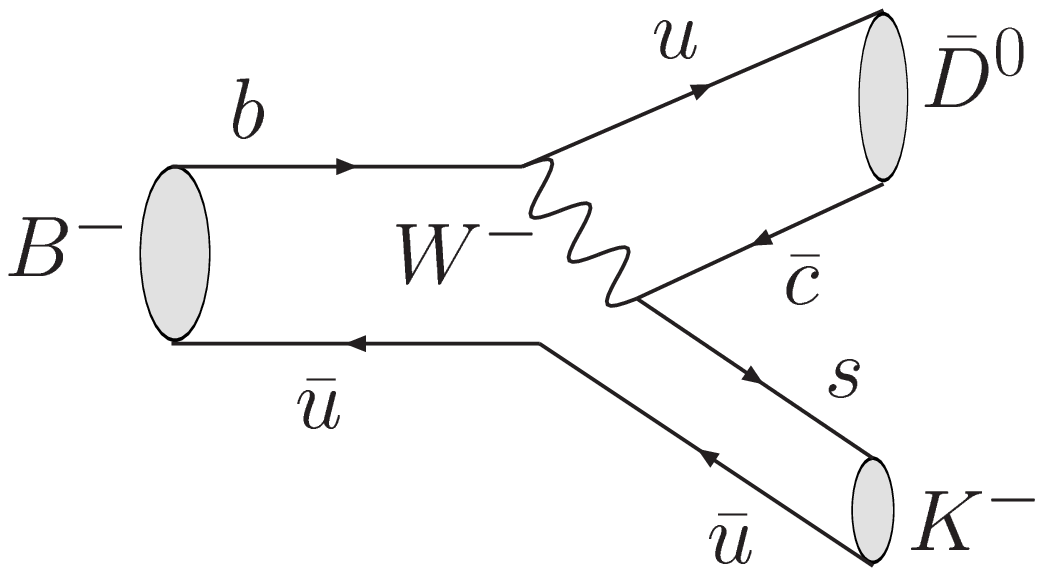,width=1.3in}}
\caption{The leading decay diagrams for 
\mbox{$B^-\to D^{0}K^{-}$ and $B^-\to \overline{D}{}^{0}K^{-}$.}}
\label{fig-btodk}
\end{figure}

Interference between competing decay amplitudes renders the relative
phase $\gamma$ observable.  The relevant processes for measuring
$\gamma$ involve interference between two tree-level diagrams, as in
Fig.~\ref{fig-btodk}, where the $D^{0}$ and $\overline{D}{}^{0}$ decay
to a common final state.  The related modes $B^-\to D^{*0}K^-$ and
$B^-\to D^0K^{*-}$ are also used.  The amplitude ratio can be
expressed as
\begin{equation}
\frac{A(B^-\to \overline{D}{}^0 K^-)}{A(B^-\to D^0 K^-)}
= r_B e^{i\delta_B} e^{-i\gamma}
\end{equation}
where $r_B$ and $\delta_B$ are the ratio of the magnitudes and the
strong phase difference of the contributing amplitudes.  In contrast
to $\gamma$, these latter parameters are specific to each $B$ decay
mode used.  These additional parameters can be determined from
observables in $B$ and $D$ decays.  The parameter $r_B$ plays an
important role in determining the experimental sensitivity to
$\gamma$, as small values render the measurements very challenging.

\subsection{Strategies for exploiting interference}
There are several ways of obtaining the same final state from
$D^0$ and $\overline{D}{}^0$ decays:
\begin{description}
\item[GLW\cite{GLW}] Choose CP eigenstates of the $D^0$ decay, e.g.
$D^0\to K^0_s\pi^0$ (CP-odd), $D^0\to \pi^+\pi^-$ (CP-even);
\item[ADS\cite{ADS}] Use doubly Cabibbo-suppressed decays (DCSD), 
e.g. $D^0\to K^+\pi^-$;
\item[GGSZ/Belle\cite{GGSZ}] Examine the $B^-$ and $B^+$
Dalitz plots for 3-body flavor-neutral
decays like $D^0\to K^0_s\pi^+\pi^-$.
\end{description}
The last method includes regions dominated by two-body decays to CP
eigenstates (e.g. $K^0_s\rho^0$) and to DCSD decay modes
(e.g. $K^{*+}\pi^-$).

\subsection{Measurements using charged $\boldmath{B}$ decays}

Both \babar\ and Belle have used each of the methods mentioned above.  The
decay $B^-\to D^{(*)0} K^{(*)-}$ is reconstructed using kinematic information,
vertex constraints and particle identification.  Particle
identification information and the difference $\Delta E$ between the
known $B$ energy in the center-of-mass frame and the energy
reconstructed from the decay products provide discrimination between
$B^-\to D^{(*)0} K^-$ and the more copious decays $B^-\to D^{(*)0}\pi^-$.
The transitions $D^{*0}\to D^0\pi^0$ and $D^{*0}\to D^0\gamma$ are
both reconstructed.

\subsubsection{GLW method}

The experiments use
$B^-\to D^{(*)}_{CP}K^{(*)-}$ decays
to measure the quantities
\begin{eqnarray}
\mathcal{A}_{\pm} &=& \frac
{\Gamma(B^-\to D_\pm K^-)-\Gamma(B^+\to D_\pm K^+)}
{\Gamma(B^-\to D_\pm K^-)+\Gamma(B^+\to D_\pm K^+)} \nonumber \\
&=& \frac
{\pm 2 r_B \sin\delta_B\sin\gamma}
{1+r_B^2\pm 2r_B\cos\delta_B\cos\gamma}
\end{eqnarray}
and
\begin{eqnarray}
\mathcal{R}_{\pm} &=& 2\frac
{\Gamma(B^-\to D_\pm K^-)+\Gamma(B^+\to D_\pm K^+)}
{\Gamma(B^-\to D^0K^-)+\Gamma(B^+\to \overline{D}{}^0K^+)}\nonumber \\
&=& {1+r_B^2\pm 2r_B\cos\delta_B\cos\gamma}
\end{eqnarray}
where the subscripts ($\pm$) indicate CP-even or CP-odd final states.
Solving these equations for the three unknowns results in up to an
8-fold ambiguity on $\gamma$.  The ability to determine $r_B$ and
$\gamma$ independently is limited due to the nature of the dependence
of the observables on these quantities.

Belle\cite{BelleGLW} and \babar\cite{BaBarGLW} 
each have $\sim 100$ events per CP eigenvalue in
$D^0K^-$ and less in $D^{*0}K^-$ and $D^0K^{*-}$, so the statistical
sensitivity is still modest.  A recent compilation of the experimental
results is available from HFAG.\cite{HFAG-CKM}  The implications on
the determination of $\gamma$ are best addressed in a global approach
that considers the GLW-based measurements in conjunction with the other
methods, and will be given later.

\subsubsection{ADS method}

The competing amplitudes in this case consist of a combination of
a favored $B$ decay and a suppressed $D$ decay or vice-versa,
e.g. $B^-\to D^0K^-$ with $D^0\to K^+\pi^-$ and
$B^-\to \overline{D}{}^0K^-$ with $\overline{D}{}^0\to K^+\pi^-$.
This can lead to large CP asymmetries, but results in small product
branching fractions.  The ratio of amplitudes also depends on
additional input from $D$ decays, namely the ratio $r_D$ and
phase difference $\delta_D$ between the suppressed and favored
$D$ decays to the specified final state.
The observables in this case are the asymmetry between $B^-$ and
$B^+$ and the ratio of suppressed to favored $B$ modes:
\begin{eqnarray}
\mathcal{A} &=& \frac
{\Gamma(B^-\to \notD_F K^-)-
 \Gamma(B^+\to \notDbar_F K^+)}
{\Gamma(B^-\to \notD_F K^-)+
 \Gamma(B^+\to \notDbar_F K^+)}\nonumber \\
&=& \frac
{2 r_B r_D \sin(\delta_B+\delta_D)\sin\gamma}{\mathcal{R}}\\
\mathcal{R} &=& \frac
{\Gamma(B^-\to \notD_F K^-)+
 \Gamma(B^+\to \notDbar_F K^+)}
{\Gamma(B^-\to D_F K^-)+
 \Gamma(B^+\to \overline{D}{}_F K^+)}\nonumber \\
=&r_D^2&+r_B^2+ 2r_Br_D\cos(\delta_B+\delta_D)\cos\gamma
\end{eqnarray}
where $D_F$ is a favored $D$ decay (e.g. $K^-\pi^+$) and
\mbox{$\notD_F$} is a dis-favored $D$ decay (e.g. $K^+\pi^-$).
The amplitude ratio $r_D$ is determined in charm decays, so
$\mathcal{R}$ has good sensitivity to $r_B$.

\babar\cite{BaBarADS} and Belle\cite{BelleADS} 
have analyzed the $B\to DK$ mode with $D$ decays to
$K^-\pi^+$, resulting in the average value $\mathcal{R}=0.006\pm 0.006$,
from which an upper limit on $r_B$ can be set.  \babar\ has also measured
$B\to D^*K$ and $B\to DK^*$, and has measured the $B\to DK$ mode
with $D\to K^-\pi^+\pi^0$ decays;\cite{BaBarADSKpipiz} 
in all cases the signal in the
suppressed mode is consistent with zero, and no asymmetry measurements
are yet possible.  The numerical results are summarized in the HFAG
compilation.\cite{HFAG-CKM}

\subsubsection{Dalitz analyses}

Three-body $D$ decays provide an opportunity to study the CP asymmetry
as a function of location in the Dalitz plot.  The three-body modes
studied consist of a neutral particle and two charged particles, e.g.
$K^0_s\pi^+\pi^-$.  The total amplitude for a given
set of mass squared values, denoted by $m_+^2$ and $m_-^2$ according
to the charge of the two-particle combination, is the sum of
the favored decay ($B^-\to D^0 K^-$) amplitude plus a
contribution from the suppressed decay ($B^-\to \overline{D}{}^0 K^-$):
\begin{eqnarray*}
A_\pm = f(m_+^2,m_-^2) + r_B e^{i(\delta_B \pm \gamma)} f(m_-^2,m_+^2).
\end{eqnarray*}
Here $f(m_+^2,m_-^2)$ denotes the amplitude for the $D^0$ decay.  In the
absence of CP violation in $D$ decays, the amplitude for the
$\overline{D}{}^0$ decay is $f(m_-^2,m_+^2)$.  A flavor-tagged
sample of $D$ decays (e.g. from $D^{*+}\to D^0\pi^+$ transitions)
is used to determine $f$.  The information contained in the
Dalitz plot removes all
ambiguities in the determination of $\gamma$ apart from the reflection
$(\gamma,\delta_B)\to(\gamma+\pi,\delta_B+\pi)$.

\begin{figure}[tb]
\centerline{\psfig{file=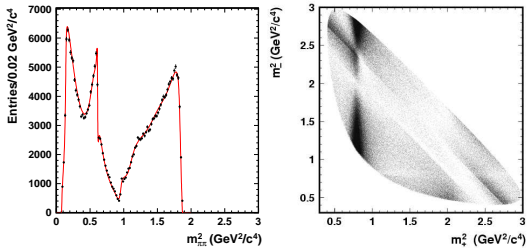,width=2.8in}}
\caption{Dalitz plot of tagged $D^0\to K_s^0\pi^+\pi^-$ from Belle
(right) and a projection onto $m^2_{\pi\pi}$ of an isobar model fit (left).}
\label{fig-DalitzModel}
\end{figure}

Belle\cite{BelleDalitz} and \babar\cite{BaBarDalitz} have analyzed
samples of $\sim 4\cdot 10^5$ $D^{*+}\to [K^0_s\pi^+\pi^-]_D\pi^+$
decays (see Fig.~\ref{fig-DalitzModel}).  The Belle (\babar) isobar
fit includes 15 (16) Breit-Wigner amplitudes plus a non-resonant term.
The main contributing resonances are $K^{*+}(892)\pi^-$,
$K^0_s\rho^0$, $K^{*0}(1430)\pi^-$, $K^{*-}(892)\pi^+$ and the
non-resonant component.  Despite the excellent qualitative description
of the data provided by the fit, the model uncertainty is a
significant source of systematic error on $\gamma$, so improved decay
modeling (e.g. a K-matrix formulation) are under investigation.  The
feasibility of a model-independent approach\cite{GGSZ} that makes use
of CP-tagged $D^0$ mesons (as could be studied at the 
$\psi(3770)\to D\overline{D}$) was studied recently.\cite{Bondar}

\begin{figure}[tb]
\centerline{\psfig{file=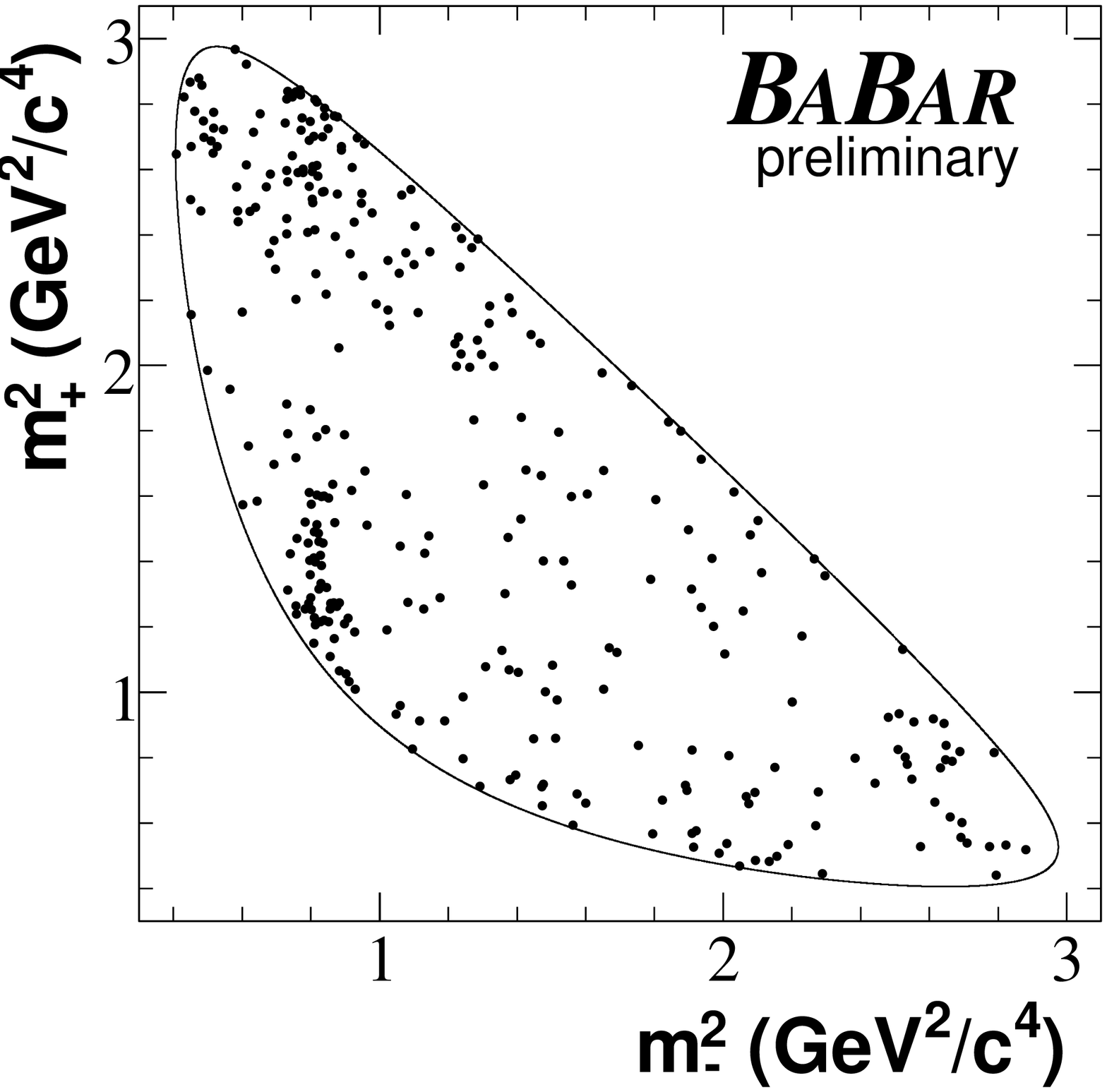,width=1.4in,bbllx=0bp,bblly=0bp,bburx=567bp,bbury=567bp,clip=}
\psfig{file=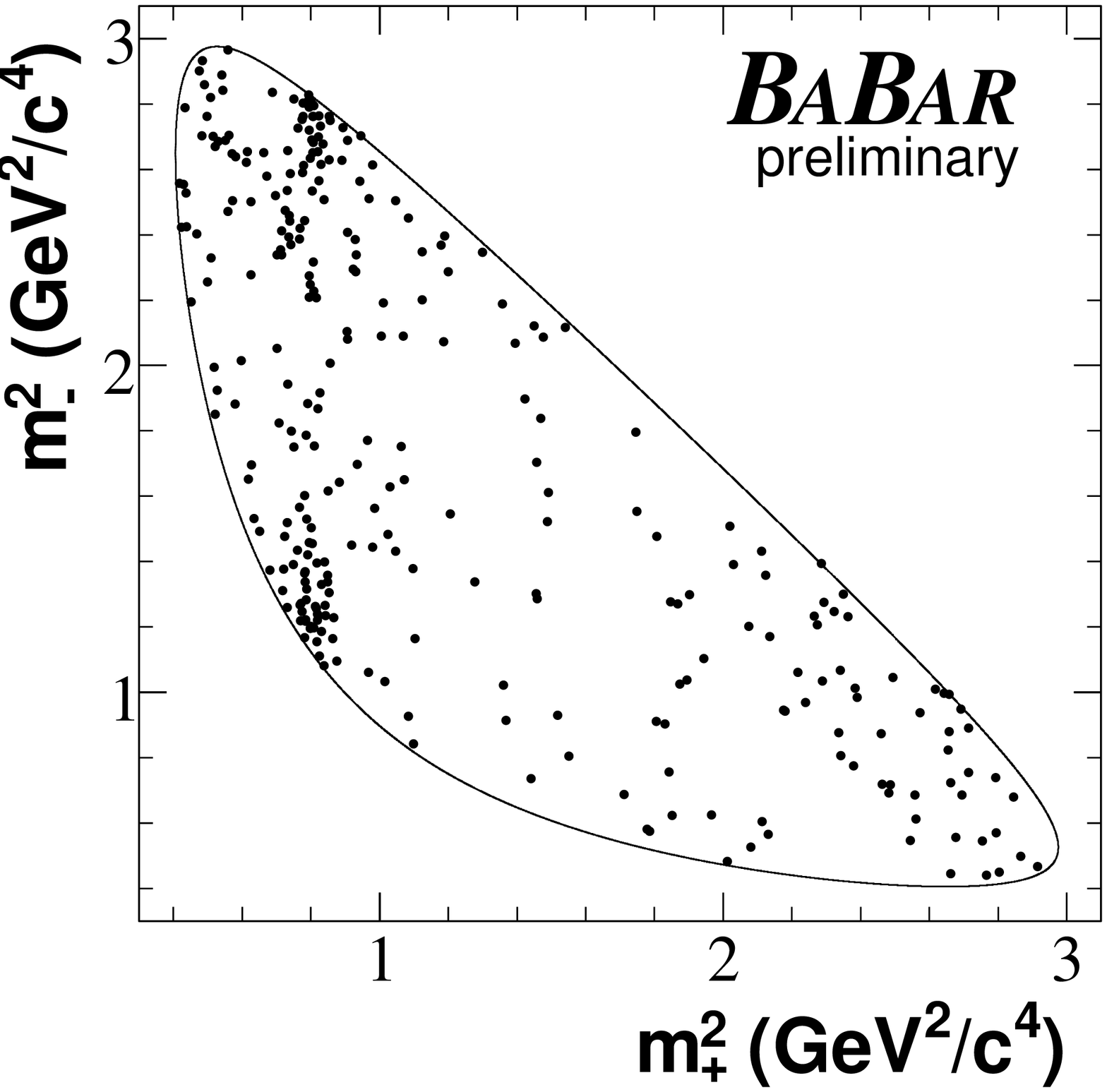,width=1.4in,bbllx=0bp,bblly=0bp,bburx=567bp,bbury=567bp,clip=}}
\caption{Dalitz plots for $B^-\to [K^0_s\pi^+\pi^-]K^-$ (left) and
$B^+\to [K^0_s\pi^+\pi^-]K^+$ (right) from \babar.}
\label{fig-BaBarDalitzDK}
\end{figure}

The Dalitz plots for $B^-$ decays into $DK^-$, $D^*K^-$ and $DK^{*-}$,
plus their charge conjugates, have been analyzed by \babar\ and Belle.
Figure~\ref{fig-BaBarDalitzDK} shows the $DK$ plots from \babar.  The
experiments extract, for each mode, contours in the $x_\pm$-$y_\pm$
plane, where $x_\pm=r_B\sin(\delta_B\pm \gamma)$ and
$y_\pm=r_B\cos(\delta_B\pm \gamma)$; these variables are uncorrelated,
in contrast to $r_B$, $\delta_B$ and $\gamma$.  The recent
measurements improve the accuracies on $x_\pm$ and $y_\pm$
significantly.\cite{HFAG-CKM} The uncertainty on $\gamma$, however, is
strongly sensitive to the value of $r_B$.  The latest measurements
favor a smaller $r_B$, and the uncertainty on $\gamma$ has increased
as a result.

Determinations of $\gamma$ from the existing measurements (GLW,
ADS and Dalitz) have been done by the UTfit collaboration,\cite{UTfit} who
find $\gamma=(82\pm 20)^\circ$, and by the CKMfitter collaboration\cite{CKMfitter},
who find $\gamma=(60^{\,+38}_{\,-24})^\circ$.  The differences arise from
a different treatment of systematic errors and different statistical
methods.  Both ranges are consistent with the values, 
$65^\circ$ and $59^\circ$, based on the respective global CKM fits.
Further progress on $\gamma$ requires larger data sets, and predictions
for the uncertainty remain uncertain due to the dependence on the 
$r_B$ values for the contributing decays, which are still not well known.

\subsection{Measurements using neutral $\boldmath{B}$ decays}

The interference between tree-level decays can also be exploited in
neutral $B$ mesons, e.g. $B^0\to D^{(*)+}\pi^-$ and
$B^0\to D^{(*)-}\pi^+$, which can interfere due to
$B\overline{B}$ mixing.  The effect is thus decay time dependent, 
and the ratio of suppressed to favored amplitudes is
\begin{equation}
\frac{A_{\mathrm{sup}}}{A_{\mathrm{fav}}}
= r_B e^{i\delta_B} e^{-i(2\beta+\gamma)} .
\end{equation}
The branching fraction is large, but $r_B$ is small ($<2\%$).  The B
factory results\cite{sin2bg} show a non-zero CP asymmetry at the
$3\sigma$ level when combined.  However, independent input is needed
for $r_B$ in order to place constraints on
$2\beta+\gamma$.  One method is to estimate the magnitude
of the suppressed amplitude
$B^0\to D^{(*)+}\pi^-$ from measurements of
$B^0\to D_s^{(*)+}\pi^-$ decays\cite{rDpi} and
SU(3) flavor symmetry; the uncertainty due to the latter
assumption is hard to quantify.

\babar\ has studied $\overline{B}{}^0\to D^{(*)0}K^{(*)0}$,\cite{B0D0K0}
from which a limit $r_{DK^*}<0.4$ is
set at 90\%\ c.l.  New measurements have been reported on $B^0\to
D_s^+ a_{0,2}^-$\cite{BDsa02} and $B^0\to D^0K^+\pi^-$;\cite{BDkpi}
neither of these channels looks promising for the determination of
$\gamma$.  A measurement of $B\to D^*\overline{D}{}^*$ branching
fractions and charge asymmetries\cite{BDstDst} provides input for
determining $\gamma$ from time-dependent $B^0\to D^{*+}D^{*-}$
decays; the corresponding constraints on $\gamma$ are weak at present.

\section{Summary and Outlook} 

\begin{figure}[htb]
\centerline{\psfig{file=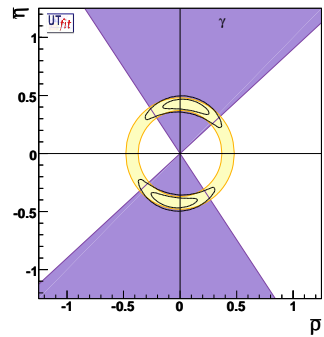,width=2.0in}}
\caption{Constraints in the $\overline{\rho}$-$\overline{\eta}$ plane
from tree-level processes from the UTfit collaboration.\protect\cite{UTfit}}
\label{fig-UTfit}
\end{figure}

The impact of the quantities reviewed here on our knowledge of CKM
parameters $\overline{\rho}$ and $\overline{\eta}$ is shown in
Fig.~\ref{fig-UTfit}.  These measurements favor a larger value for
$\sin 2\beta$ than is determined from CP asymmetry measurements.  The
discrepancy is at the level of $1.5$-$2\sigma$, and depends in detail
on the assumptions made in the global fits.  It is nevertheless
intriguing, and underscores the motivation for further improvements in
accuracy.

The B factories will continue to increase their data samples, and
expect to have $2\,$ab$^{-1}$ between them by ICHEP 2008.
This will allow for significant improvements in
$|V_{ub}|$ and $\gamma\ (\phi_3)$.  Additional improvements in the
calculations used to extract $|V_{ub}|$ and $|V_{cb}|$ from both
inclusive and exclusive semileptonic decays can be expected.  These
refinements will further restrict the space in which theories hoping
to explain the new physics we'll see at the LHC can live.

{\bf Acknowledgments}

The author would like to thank the conference organizers for their kind
hospitality, and to thank Masahiro Morii and Tim Gershon for
helpful discussions.

\end{document}